\documentclass[%
    superscriptaddress,%
    amsmath,%
    amssymb,%
    floatfix,
    twocolumn,
    ]{revtex4-1}

\usepackage{graphicx}
\usepackage{dcolumn}
\usepackage{bm}
\usepackage{tabularx}
\usepackage{xspace}

\newcommand{\ttas}{1T-TaS\textsubscript{2}\xspace}
\newcommand{\cf}{cf.\xspace}
\newcommand{\ie}{i.\,e.\xspace}
\newcommand{\ti}[1]{$\mathbf{t}_{#1}$\xspace}
\newcommand{\talt}[1]{$\mathbf{t}_{\mathrm{alt}}$\xspace}

\begin{document}
\title{Collapse of layer dimerization in the photo-induced hidden state of 1T-TaS$_2$}

\author{Quirin Stahl}
\author{Maximilian Kusch} 
\affiliation{Institut f\"ur Festk\"orper- und Materialphysik, Technische
Universit\"at Dresden, 01062 Dresden, Germany}

\author{Florian Heinsch} 
\affiliation{Institut f\"ur Festk\"orper- und Materialphysik, Technische
Universit\"at Dresden, 01062 Dresden, Germany}
\affiliation{Institute of Radiation Physics, Helmholtz-Zentrum Dresden-Rossendorf, 01328 Dresden, Germany}

\author{Gaston Garbarino}
\author{Norman Kretzschmar}
\affiliation{ESRF, The European Synchrotron, 38000 Grenoble, France
}

\author{Kerstin Hanff}
\affiliation{Institut f\"ur Experimentelle und Angewandte Physik,
  Christian-Albrechts-Universit\"at zu Kiel, 24098 Kiel, Germany}

\author{Kai Rossnagel}
\affiliation{Institut f\"ur Experimentelle und Angewandte Physik,
  Christian-Albrechts-Universit\"at zu Kiel, 24098 Kiel, Germany}
\affiliation{Ruprecht-Haensel-Labor, Christian-Albrechts-Universit\"at zu Kiel und Deutsches Elektronen-Synchrotron DESY, 24098 Kiel und 22607 Hamburg, Germany} 

\affiliation{Deutsches Elektronen-Synchrotron DESY, 22607 Hamburg, Germany}

\author{Jochen Geck}
\affiliation{Institut f\"ur Festk\"orper- und Materialphysik, Technische
Universit\"at Dresden, 01062 Dresden, Germany}
\affiliation{W\"urzburg-Dresden Cluster of Excellence ct.qmat, Technische
Universit\"at Dresden, 01062 Dresden, Germany}

\author{Tobias Ritschel} 
\affiliation{Institut f\"ur Festk\"orper- und Materialphysik, Technische
Universit\"at Dresden, 01062 Dresden, Germany}

\begin{abstract}
Photo-induced switching between 
collective  quantum states of matter 
is a fascinating rising field with exciting opportunities for novel technologies.
Presently very intensively studied examples in this regard are nanometer-thick single crystals of the layered material \ttas, where picosecond laser pulses
can trigger a fully reversible insulator-to-metal transition (IMT). This IMT is believed to be connected to the switching between metastable collective quantum states, but the microscopic nature of this so-called hidden quantum state remained largely elusive up to now. 
Here we characterize the hidden quantum state of \ttas by means of state-of-the-art x-ray diffraction and show that the laser-driven IMT
involves  a marked rearrangement of the charge and orbital order in the direction perpendicular to the TaS$_2$-layers.  
More specifically, we identify the collapse of inter-layer molecular orbital dimers as a key mechanism for this non-thermal collective transition between two truly long-range ordered electronic crystals.
\end{abstract}

\maketitle

The layered transition metal dichalcogenides (TMDs) form a vast class of
materials hosting diverse non-trivial quantum phenomena such as spin-valley polarization~\cite{Xiao2012},
Ising-superconductivity~\cite{Xi2015a} or intertwined electronic orders~\cite{Sipos2008,Ritschel2015}. All these intriguing
electronic effects
along with the natural suitability of TMDs for the preparation of quasi two-dimensional (2D) nano-sheets render them highly appealing for next-generation technologies~\cite{Radisavljevic2011,Jariwala2014,Stojchevska2014,Yoshida2015}.

\ttas is a particularly interesting and extensively studied TMD in which 
external tuning parameters like
temperature, pressure or chemical substitution span a very complex
electronic phase diagram. Apart from several charge density waves (CDWs)
this phase diagram also features
pressure-induced superconductivity 
and a so-called Mott-phase, which stands out due to its
semiconducting electronic transport properties~\cite{Wilson1975a,Sipos2008}. 

Remarkably, besides the aforementioned states that can be reached in thermal equilibrium, femto to picosecond optical or electrical pulses can trigger a non-equilibrium IMT into a previously hidden and persistent metallic CDW-state~\cite{Stojchevska2014,Vaskivskyi2015,Hollander2015}. The discovery of this so-called hidden CDW (HCDW) has sparked wide excitement 
as it might provide a new  platform for memory device applications.
Accordingly, in recent years, a significant number of experimental and theoretical
studies aimed at pinning down the microscopic mechanism of this non-equilibrium
IMT that is believed to be connected to a reorganization of the CDW-order. Although significant efforts to determine the microscopic processes underlying this novel IMT have been made~\cite{Guyader2017,Hovden2018,Ma2016,Cho2016,Cho2017,Gerasimenko2019a}, a clear picture remains elusive. 

In this article we address this open issue directly by means of high-resolution synchrotron x-ray diffraction (XRD) in combination with laser pumping. 
Our experiments enable examination of the laser-driven transition and in particular the HCDW-order in \ttas nano-sheets with great sensitivity and in all three spatial directions. 
In this way we show that ultra-short laser pulses drive a collapse of the inter-layer dimerization present in thermal equilibrium, which reveals the key physical process behind the 
laser-driven IMT of \ttas.

\begin{figure*}[htpb]
  \centering
  \includegraphics[]{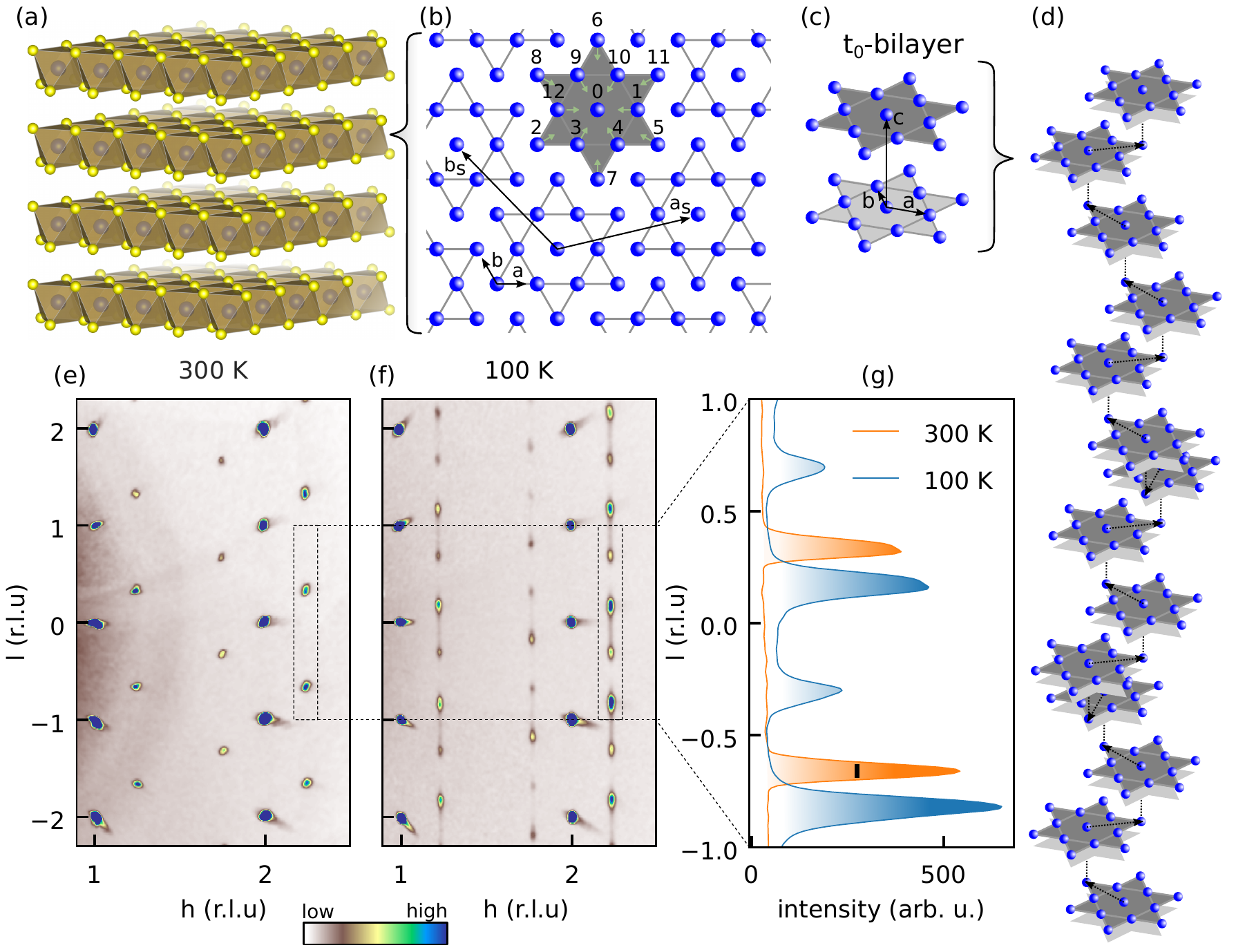}
  \caption{\textbf{Structure and CDW layer stacking in \ttas.} (a) The layered host
    structure of \ttas comprises S-Ta-S layers (Spacegroup: \textit{P}$\bar3$\textit{m}1). (b)
    The structural key feature of the C and NCCDW are  star-of-David shaped (SOD)
    clusters containing 13 Tantalum sites formed by an inward displacement (green
    arrows) of 12 Ta ions (labels 1\ldots12) towards the central Ta ion (label 0).
    The SOD themselves form a $\sqrt{13} \times\sqrt{13}$-superlattice with in-plane
    lattice vectors $a_s$ and $b_s$. (c) and (d) The partially disordered
    stacking of the CCDW at low temperatures is dominated by \ti0 bilayers (c)
    which by themselves are stacked with a vector randomly drawn from the three
    symmetry equivalent vectors \ti7, \ti8 and \ti{11} (d). The \ti0 bilayers are
    illustrated as gray SODs in (d). (e) and (f): Reciprocal space maps parallel to
    the $h0l$-plane (integration thickness perpendicular to the plane: $\delta k = 0.3$ r.l.u) for
    T=300\,K (e: NCCDW) and T=100\,K (f: CCDW). (g) Intensity profile of the superlattice
    reflections along the $l$-direction for CCDW (blue) and NCCDW (orange). The
    double-peak feature is the key fingerprint for the \ti0 bilayers present in the
    CCDW. The black bar indicates the full width at half maximum of a typical Bragg peak.}
  \label{introfig}
\end{figure*}

\section*{Results}

\subsection{CDW-states in thermal equilibrium}

At ambient pressure and at temperatures below $\approx 180$\,K a commensurate charge density wave (CCDW) develops within the layers of \ttas, which is characterized by the
formation of star-of-David (SOD) shaped clusters containing 13 Ta-sites arranged
in a $\sqrt{13}\times\sqrt{13}$ superlattice (cf. Fig~\ref{introfig}~(a,b)). 
The periodic lattice distortion (PLD) associated with a CDW has a larger periodicity than the underlying crystal lattice and becomes visible in XRD through the appearance of characteristic superlattice reflections which are typically $\gtrapprox 10^2$  times weaker than the Bragg reflections stemming from the average structure. For the CCDW these reflections occur at the commensurate q-vectors $q^\mathrm{C}_1 = (\sigma_1^\mathrm{C}, \sigma_2^\mathrm{C}, l)$ and $q^\mathrm{C}_2 = (-\sigma_2^\mathrm{C}, \sigma_1^\mathrm{C}+\sigma_2^\mathrm{C}, l)$ with $\sigma_1^\mathrm{C}=3/13$ and $\sigma_2^\mathrm{C}=1/13$~\cite{Brouwer1980,Spijkerman1997}.
The superlattice structure of the CCDW is very well ordered within the
$ab$-plane which translates  into sharp superlattice reflections in the $hk$-plane. 
However, in the direction perpendicular to
the S-Ta-S-layers, the CDW order is subject to partial disorder among the 13
possible two-layer stacking arrangements
labeled as $\mathbf{t}_0 \ldots \mathbf{t}_{12}$ corresponding to the 13 Ta-sites within one star-of-David cluster  (\cf Fig.~\ref{introfig}~(b)). 

Due to this disorder in one direction the superlattice relections form diffuse stripes along the  $l$-direction rather than sharp peaks. 
The intensity distribution along these stripes has a characteristic two-peak
structure (see Fig.~\ref{introfig}~(f,g)). A careful analysis of the diffuse-scattering reveals that the  3D arrangement of the CDW in \ttas at low temperatures is given by a stacking in which bilayers obeying the $\mathbf{t}_0$-stacking occur, as illustrated in  Fig.~\ref{introfig}~(c).
These bilayers are arranged in a stacking sequence, where the stacking vector
between two adjacent bilayers is randomly 
chosen from the three symmetry equivalent vectors \ti7, \ti8 and \ti{11}, which is schematically depicted in
Fig.~\ref{introfig}~(d)~\cite{Walker1983,Nakanishi1984,Tanda1984,Ritschel2018}.

It should be noted that merely the superlattice reflections show diffuse
scattering stripes while the Bragg-peaks remain sharp, indicating the absence of more conventional stacking faults in the average structure. 

The CCDW phase is also often referred to as the  Mott-phase,
because its unusual semiconducting transport properties are widely believed to stem from
Mott-Hubbard type electron-electron correlations~\cite{Tosatti1976}. But this paradigm
is currently under intense debate as several studies find that the CDW-stacking in the
OP-direction has a marked influence on the low-energy electronic structure. In fact, the CDW-stacking can also drive the formation of a charge excitation gap at the Fermi energy and, hence, yield the observed semiconducting transport behavior. The crucial ingredient for the stacking-induced gap-formation seems to be the occurrence of the $\mathbf{t}_0$-bilayers~\cite{Ge2010,Ritschel2018,Ritschel2015,Lee2019}. 

Increasing the temperature or applying external pressure transforms the CCDW
into a so-called nearly commensurate CDW (NCCDW), accompanied by a significant
metallization of the sample. 
The NCCDW is structurally characterized by the formation of a well-ordered network of defects -- so-called discommensurations -- within the
$\sqrt{13}\times\sqrt{13}$-superlattice of
SODs~\cite{Nakanishi1978a,Wu1989,Spijkerman1997}.
This in-plane modification of the CDW structure takes place on a length-scale of about 70\,\AA{} and is accompanied  by a marked rearrangement of the CDW order in
the out-of-plane direction. In XRD, the fingerprints of the NCCDW are given by
a slight shift of the q-vectors to $q^\mathrm{NC}_1 = (\sigma_1^\mathrm{NC}, \sigma_2^\mathrm{NC}, 1/3)$ and $q^\mathrm{NC}_2
= (-\sigma_2^\mathrm{NC}, \sigma_1^\mathrm{NC}+\sigma_2^\mathrm{NC}, -1/3)$ with $\sigma_1^\mathrm{NC}=0.248$ and $\sigma_2^\mathrm{NC}=0.068$~\cite{Spijkerman1997}, along with the appearance of strong
higher-order superlattice reflections owing to the domain-like structure imposed by the discommensurations. Instead of the diffuse stripes along the $l$-direction
which occur for the CCDW, the superlattice reflections of the NCCDW are sharp
and centered at  $l_{NC} = \pm 1/3$, indicating a well-ordered CDW stacking with a periodicity of three times the interlayer distance and the absence of the $\mathbf{t}_0$ bilayers (\cf Figure~\ref{introfig}~(e,g)). Although several scenarios for the microscopic mechanism of the IMT associated with the NCCDW-CCDW transition are currently being discussed, there is increasing theoretical and experimental evidence that exactly this rearrangement of the
CDW stacking plays a crucial role~\cite{Ritschel2018,Lee2019,Park2019}.

Since the photoinduced HCDW is also accompanied by a metallization of
\ttas{}, the question arises whether or not  the CDW stacking in the OP-direction is affected by this transition, too. 
In order to address exactly this issue, we performed XRD
on thin exfoliated samples of \ttas in which we induced the HCDW
by means of a single 1.6 ps pulse from a Ti-sapphire laser. The results of these
measurements will be presented in the following section.

\subsection{Photo-induced metastable CDW-state}

The presented experiments have been performed on thin flakes exfoliated from high quality \ttas single crystals 
at the beamline ID09b at the ESRF (see methods section for details).

\begin{figure*}
  \includegraphics[]{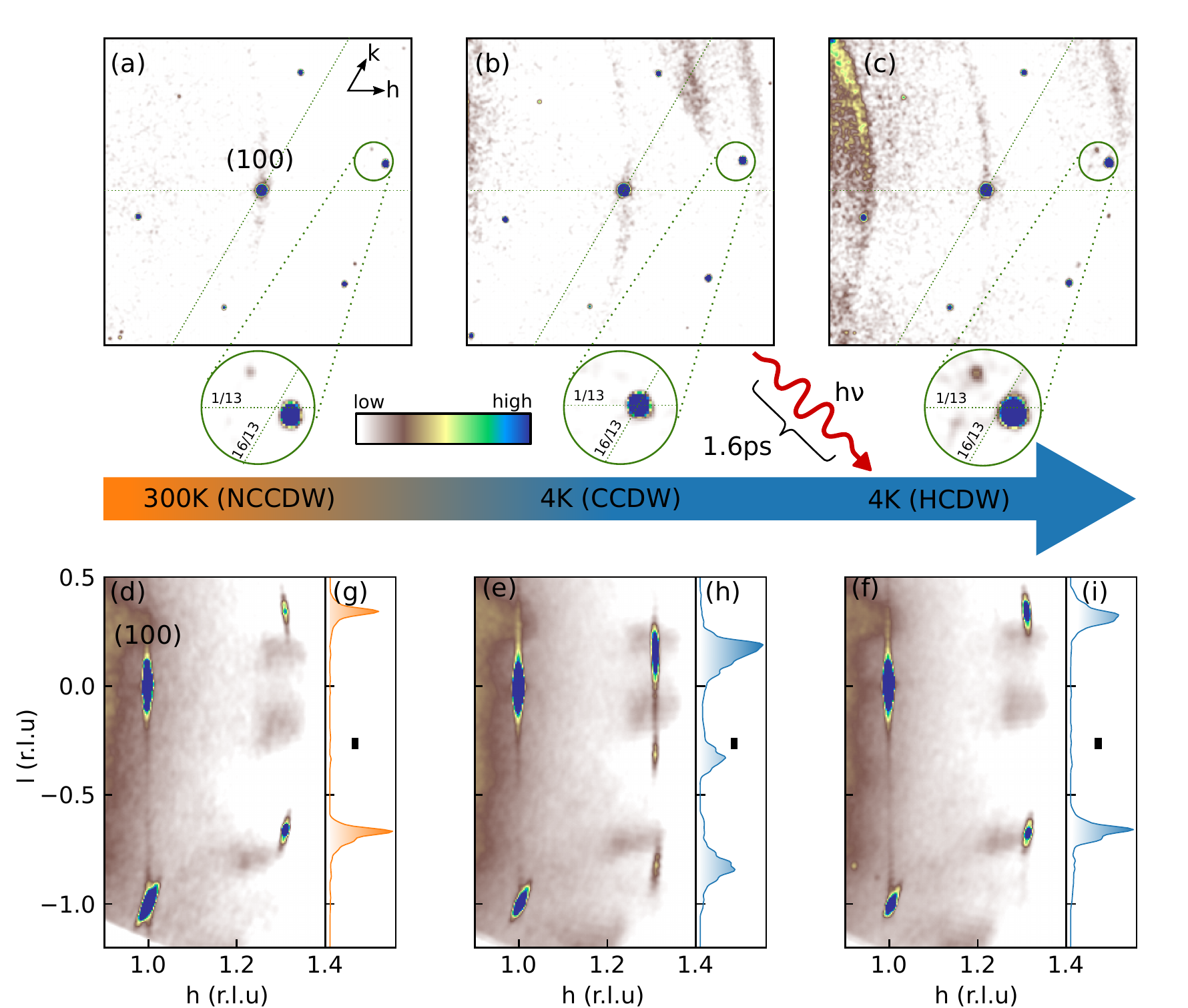}
  \caption{\textbf{Photo-induced changes of the XRD from an thin flake of \ttas.} 
    Reciprocal space maps parallel to the $hk0$-plane (integration thickness perpendicular to the plane: $\delta l = 2/3$\,r.l.u.) for the NCCDW at room temperature (a), the CCDW at $T=4$\,K (b) and the photo-induced HCDW at the same temperature $T=4$\,K. 
    The characteristic occurrence of strong higher order superlattice reflections due to the in-plane discommensurations network is clearly observable for the NCCDW (a) and the HCDW (c).
    Reciprocal space maps parallel to the $h0l$-plane (integration thickness perpendicular to the plane: $\delta k=0.3$\,r.l.u.) are shown in (d-f) for NCCDW, CCDW and HCDW, respectively. 
    The extracted intensity distribution of the superlattice reflections along
    the $l$-direction clearly shows the well defined sharp spots for the NCCDW
    (g), the characteristic diffuse double peak feature for the CCDW (h) and the
    vanishing of the diffuse scattering for the photo-induced HCDW. The black
    bars in (g),(h) and (i) indicate the full width at half maximum of a typical
    Bragg peak.) 
}
\label{fig:Reciprocal}
\end{figure*}

In Fig.~\ref{fig:Reciprocal} we summarize the key results of these measurements by showing reconstructed reciprocal space maps for a switching cycle 
\[\mathrm{(NCCDW)\xrightarrow{cooling}(CCDW)\xrightarrow{laser~pulse}(HCDW)}\]
of a thin flake \ttas sample. 
Figs.~\ref{fig:Reciprocal}~(a-c) illustrate the changes of the 
in-plane components of the superlattice reflections by projecting the diffraction pattern along the $l$-direction within a range of $-1/3 < l < 1/3$ onto the $hk0$-plane,
\ie the plane parallel to the S-Ta-S layers. In order to visualize the changes
of the out-of-plane component and the peak profile of superlattice reflections in the
OP-direction, we project the diffraction pattern onto the $h0l$-plane within a $k$-range of $-0.1 < k < 0.1$ in Fig.~\ref{fig:Reciprocal}~(e-f).

Starting at room temperature, we observe the characteristic higher order
superlattice reflections in the projection onto the $hk$-plane, which indicates the presence of ordered discommensurations in the NCCDW (see Fig.~\ref{fig:Reciprocal}~(a)). The IP-parameters of the $q$-vector $\sigma_1^\mathrm{NC}=0.245\pm0.002$ and $\sigma_2^\mathrm{NC}=0.067\pm0.002$ are in excellent agreement with the values obtained for single crystal
bulk samples~\cite{Spijkerman1997,Ritschel2013}.
Moreover, for these thin flakes the NCCDW superlattice relections are sharp along the $l$-direction and appear at $l=\pm1/3$ indicating the tripling of the unit cell in the  out-of-plane direction (Fig.~\ref{fig:Reciprocal}~(g)).
After  cooling the sample to
4\,K  the higher order superlattice reflections vanish as the in-plane components of the $q$-vector assume the commensurate values $\sigma_1^\mathrm{C} = 3/13$ and $\sigma_2^\mathrm{C}=1/13$. 
(see Fig.~\ref{fig:Reciprocal}~(b)). As can be observed in Fig.~\ref{fig:Reciprocal}~(e,h), the superlattice peaks exhibit characteristic diffuse stripes with a double peak structure along the $l$-direction. This is also found for bulk single crystals and indicates the peculiar CDW-stacking of the CCDW described above. 
We can therefore conclude that our thin exfoliated flakes of \ttas and bulk single crystals of \ttas develop the same in- and out-of-plane CDW-stacking.

Having established that the diffraction pattern unambiguously reveals the
formation of the CCDW at low temperatures, 
we continued by photoexcitation of the system. At a nominal temperature of 4\,K
we excited the sample with a single 1.6\,ps long laser pulse with a wavelength of 800\,nm
and a pulse energy of 0.18\,$\mu$J which corresponds to a fluence in the range of 1\,mJ/cm$^2$.
According to previous reports, this procedure creates the HCDW~\cite{Stojchevska2014}. Indeed, in response to the
photo-excitation the diffraction pattern changes very clearly and the higher order superlattice peaks in the projection onto the $hk$-plane appear again. This is qualitatively similar to the NCCDW at room temperature, yet with somewhat different in-plane components of the q-vector, namely
$\sigma^\mathrm{H}_1 = 0.243\pm0.002$ and $\sigma^\mathrm{H}_2 = 0.070\pm0.003$ and, hence, a smaller discommenurability $\delta = |q^\parallel_{H} - q^\parallel_C|$ as compared to the NCCDW (see also Table~\ref{tab:qIP} for a comparison of the in-plane $q$-vector parameters). Even more striking is that the peak profile in the
$l$-direction becomes sharp again and shifts back to $l=\pm1/3$. In other words, the double peak structure, characteristic for the \ti0-bilayers in the CCDW, disappears upon laser pumping. This strong effect of the laser pump pulse can be clearly observed in Figs.~\ref{fig:Reciprocal}~(e,f) and Figs.~\ref{fig:Reciprocal}~(h,i).

For all three states the superlattice peak widths within the plane are always comparable to the width of the Bragg peaks, which indicates that the IP-superstructure remains long-range
ordered. Note that the XRD data for the HCDW has been taken several seconds after the laser pulse hit the sample, which means that the sample temperature is well-defined at $T=4$\,K. The fact that the CCDW would be the thermodynamic stable state under these
conditions points to the metastable nature of the HCDW. Furthermore, the resolution limited peak width observed here, shows that the non-equilibrium process driven by the 1.6~ps laser pulse results in a truly long-ranged ordered HCDW-state.

\begin{table*}
  \begin{ruledtabular}
    \begin{tabular}{cccccc}
      state  & $\sigma_1$ (r.l.u.)       & $\sigma_2$ (r.l.u)       & $\delta$ (10$^{-2}$ r.l.u)   & $\phi$ (deg) & $|q|$ (r.l.u)\\
      \hline
      C & $3/13$             & $1/13$             & 0 & 13.898& 0.277\\
      H & $0.243  \pm 0.002$ & $0.070  \pm 0.003$ & $1.06  \pm 0.16$ & $12.3\pm0.3$& $0.284\pm0.002$\\
      NC   & $0.245\pm0.002$    & $0.067\pm0.002$    & $1.32\pm0.16$ &$11.9\pm0.3$& $0.284\pm0.002$ \\
    \end{tabular}
    \caption{\textbf{Parameters of the in-plane $q$-vector for CCDW, HCDW and NCCDW.}
    $\sigma_1$ and $\sigma_2$ are the in-plane components. $\delta$ is the
  incommensurability $\delta = |q_H^\parallel - q_C^\parallel|$. $\phi$ is the
angle of the in-plane $q$-vector with respect to the reciprocal $a^*$ axis and $|q|$ is
the length of the in-plane $q$-vector.}
    \label{tab:qIP}
  \end{ruledtabular}
\end{table*}

\subsection{Temperature driven relaxation}

\begin{figure}[htpb]
  \centering
  \includegraphics[]{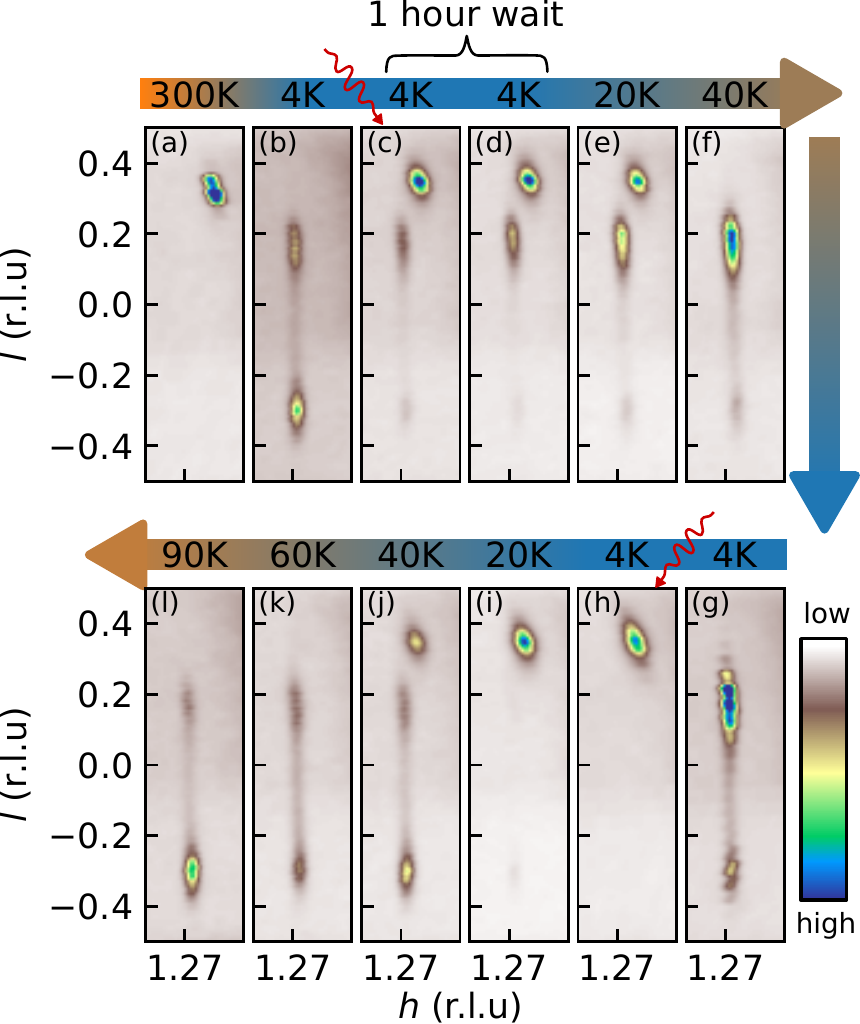}
  \caption{\textbf{Thermally driven HCDW-CCDW transition for two successive laser
  pulse/heating cycles.} (a) The NCCDW superstructure peak at room temperature. (b) The diffuse CCDW double
  peak at 4\,K. (c): The peak $l$-profile after the first photo-excitation at 3\,K. Coexistence of
    CCDW peak and HCDW peak indicating only partial switching. (d) Same as (c)
    after waiting for 1 hour, demonstrating the persistence of the HCDW. (e-f) Peak $l$-profiles for heating up to 40\,K where the HCDW peak vanishes. 
    (g) CCDW peak $l$-profile after cooling to 4\,K again. (h) Peak $l$-profile of the HCDW after a second laser
    pulse with higher fluence resulting in a more complete switching. (i-l) Second warming cycle up to 90\,K.
    The CCDW peaks reappear at around 40\,K, in (j).}
  \label{cycle}
\end{figure}

In Fig.~\ref{cycle} we show the evolution of the superstructure peaks for two
successive laser pulse/heating cycles on the same sample.  As can be clearly
seen in Fig.~\ref{cycle}~(e,f) and (j,k) the sharp peak associated with the HCDW
disappears in a temperature range around 40\,K and the system flips back to the
CCDW.  This is indeed a very characteristic property of the HCDW which was found
in previous resistivity measurements~\cite{Stojchevska2014}. Notably, this temperature is significantly lower than the
thermally driven  CCDW-NCCDW transition, at about
200\,K. Our observation of the same behavior provides strong
evidence that the photo-induced state created in our experiment is indeed the
hidden state which had been reported previously.

We also mention that for some samples we observed an incomplete photo-induced 
switching from the CCDW to the NCCDW, resulting in the simultaneous occurrence of both types of corresponding superlattice reflections (\cf Fig.~\ref{cycle}~(c)). This observation, which we attribute to a slight mismatch of sample size, laser spot size and x-ray spot size,
demonstrates that both types of long-range orders -- the CCDW and the HCDW -- can
macroscopically coexist in the same sample.

\subsection{Influence of laser fluence}

We also studied the effect of the laser pulse fluence on the induced HCDW by
increasing the laser power: 
Most importantly, the sharpening of the superlattice reflection along the $l$-direction
remains  unaffected by changing the laser pulse fluence which means that the
breakdown of the \ti0 bilayers fully takes place as soon as the switching threshold is
reached. Further we found evidence that the in-plane $q$-vector of the HCDW
slightly shifts towards the position of the room temperature NCCDW in-plane $q$-vector 
with increasing laser fluence. This trend is consistent with pump-probe 
electron diffraction experiments at higher
temperatures~\cite{Han2015}.

\section{Discussion}%
\label{sec:discussion}

We now turn the discussion towards the origin of the metallicity of the HCDW in
the light of our  observations. Our data reveal that the metallization in the
HCDW coincides with a reordering of the CDW in the out-of-plane direction, as
well as the formation of disommensurations. Recent studies based on density
functional theory (DFT) provide strong evidence that the changes in the
out-of-plane direction indeed play an important role for the IMT
~\cite{Lee2019,Ritschel2018,Ritschel2015}. 
More specifically, these studies identify a key feature responsible for the insulating properties of the CCDW, namely 
the formation of \ti0 bilayers characteristic for the  CCDW stacking along the out-of-plane direction. 
Our experimental data show that these bilayers indeed collapse in response to the picosecond
laser pulse as can be unambiguously seen from the vanishing double peak
structure in the intensity distribution of the superlattice reflections along the $l$-direction (\cf Figure~\ref{fig:Reciprocal}~(d-i)).
The same breakdown of the \ti0 bilayers also happens as a function of
temperature or pressure at the CCDW-NCCDW transition, which is also accompanied
by a marked change of the electrical resistivity. We therefore conclude that the
rearrangement of the CDW structure in the out-of-plane direction involving the
collapse of the \ti0-bilayers lies at the core of the photo-induced
IMT. 

The formation of the \ti0-bilayers is closely related to the hybridization 
between the orbitally ordered CDW layers, where each SOD represents one
quasi-molecular orbital. In other words, the CCDW phase is characterized by
a dimerization of SODs in the out-of-plane direction, i.e., the formation of
interlayer dimer-bonds~\cite{Ritschel2015,Ritschel2018,Lee2019}.  According to
our observations, photo-excitation breaks exactly these interlayer
dimers. Since this primarily is an electronic process involving orbital degrees
of freedom, it is likely to proceed on ultra-fast electronic time scales, as
observed in previous time-resolved
experiments~\cite{Perfetti2008,Hellmann2010a,Hellmann2012,Ligges2018}.

We would like to stress that the HCDW structure in our experiments is always long range
ordered throughout the thin sample. In contrast, recent STM experiments on the
HCDW show a strongly disordered network of discommensurations at the
surface~\cite{Gerasimenko2019a}. The sharp superlattice reflections and the
absence of significant diffuse scattering in our bulk sensitive XRD measurements
of the HCDW instead imply that the discommensurations form a highly ordered
network similar to the NCCDW at room temperature. This comparison therefore implies that the photo-induced processes at the surface and in the bulk are different, which is certainly relevant with respect to the down-scaling of the sample thickness for possible technological applications.

Regarding the photo-induced IMT in the bulk, our data reveal a marked similarity between the HCDW and the NCCDW. 
In spite of these clear similarities, it has been argued previously that the
NCCDW and HCDW are distinct states because they differ in their electrical
resistivity and the frequency of the amplitude mode of the
CDW~\cite{Stojchevska2014,Vaskivskyi2015}. Indeed, besides the aforementioned
similarities of the HCDW and the NCCDW, the present XRD-study uncovers also an
important difference between these two states, namely a different density of
discommensurations. Our measurements show that the in-plane incommensurability
$\delta$ of the HCDW $q$-vector is  smaller than that of the room temperature
NCCDW (see Table~\ref{tab:qIP}). This implies a smaller density of
discommensurations in the HCDW as compared to the room temperature
NCCDW~\cite{Nakanishi1978a,Ritschel2013}.
It is likely that salient features of the discommensurations network, such as the concentration of discommensurations, alter the transport properties as well as the frequency of the amplitude mode. 
Nevertheless, it is also known that the density of discommensurations 
changes as a function of temperature within the NCCDW and one could speculate 
that the HCDW is indeed a quenched form of the NCCDW, created by the transient
heating of the laser pulse and the abrupt cooling afterwards. In that spirit,
HCDW and NCCDW would be two instances of the same phase, distinguished by details
of their discommensurations network.

In conclusion, we employed synchrotron XRD in combination with picosecond
laser-pumping to uncover the structure of the photo-induced HCDW in \ttas
nanosheets. Our measurements indicate a key mechanism of the photo-induced IMT: The
collapse of interlayer dimers, which are a characteristic feature of the CCDW
and which have been proposed to be the origin of
its insulating properties by recent theoretical studies ~\cite{Lee2019,Ritschel2018}. 
Thus, the present study implies that the reorganization of the orbitally ordered layers
in the out-of-plane direction, rather than in the in-plane direction lies at the core of the marked transport anomalies in \ttas. On a more general level, the case at hand illustrates in a striking manner that out-of-plane correlations between the layers in Van der Waals materials can play a crucial role for their physical properties. This is not a mere complication but provides a unique way to control the electronic properties of such materials in the few layer limit. 

A very intriguing and surprising feature of the studied case is that a single
picosecond laser pulse can trigger a transition between two very complex and
truly long-range ordered electronic states, which are the endpoints of the
photo-induced IMT. How this transition proceeds on atomic length and electronic
time scales still remains an open and very fascinating puzzle. A complete
understanding of this unusual collective electronic phenomenon will
require further time-resolved experiments.

\section{Methods}
\subsection{Sample preparation}
The high-quality 1T-TaS$_2$ bulk single crystals used for the present study were
were grown from high purity elements by chemical vapour transport using iodine as transport agent.
Thin films with a thickness of about 50\,nm and typical lateral dimensions up to
200\,$\mu$m were prepared by mechanical exfoliation. The exfoliated samples were
transferred onto standard TEM silicon nitride windows with 200\,nm thickness.

\subsection{XRD measurements and laser excitation}

Bulk single crystal XRD data at room temperature and 100 K were measured
at a  Bruker APEXII diffractometer.

XRD experiments on the exfoliated \ttas samples were performed at the beamline ID09 of the European Synchrotron Radiation Facility.  
The TEM silicon nitride window with the exfoliated \ttas samples  were mounted on
the cold finger of a specially designed continuous He-flow cryostat with Mylar windows which are transparent for both the optical laser radiation and the 18 keV x-ray radiation. The cryostat was attached to
a single rotation axis. 
Photo-excitation of the exfoliated \ttas samples was achieved by means of a single pulse from an optical Ti-sapphire laser with a wavelength of 800\,nm and a pulse duration of 1.6 ps.  
The laser spot size on the sample was about 400 $\mu$m (FWHM) while the x-ray
beam spot size was about $60\times100$ $\mu$m$^2$.
XRD single crystal datasets from the samples were collected using a  Rayonix MX170 detector in transmission geometry.
All data sets contain 120 frames with 0.5 deg scan width over a sample rotation
of 60 deg. The diffraction images were transformed into reciprocal space using
the CrysAlis Pro software package~\cite{CrysAlisPRO}. Reciprocal space maps as
shown in Figure~\ref{introfig}~and~\ref{fig:Reciprocal} were produced using the
Python packages numpy, matplotlib, and fabio.

\section{Acknowledgements}
This research  has been supported by the Deutsche Forschungsgemeinschaft through the SFB 1143, the Würzburg-Dresden Cluster of Excellence EXC 2147 (ct.qmat), the Graduate School GRK 1621, and Grant No.  RI 2908/1-1. We also gratefully acknowledge  the  support provided  by the DRESDEN-concept alliance of research institutions, thank T. Woike and M. Wulff for helpful discussions and the ESRF for providing  the beamtime at ID09.


\begin{thebibliography}{10}
\expandafter\ifx\csname url\endcsname\relax
  \def\url#1{\texttt{#1}}\fi
\expandafter\ifx\csname urlprefix\endcsname\relax\def\urlprefix{URL }\fi
\providecommand{\bibinfo}[2]{#2}
\providecommand{\eprint}[2][]{\url{#2}}

\bibitem{Xiao2012}
\bibinfo{author}{Xiao, D.}, \bibinfo{author}{Liu, G.-B.},
  \bibinfo{author}{Feng, W.}, \bibinfo{author}{Xu, X.} \& \bibinfo{author}{Yao,
  W.}
\newblock \bibinfo{title}{Coupled spin and valley physics in monolayers of
  ${\mathrm{mos}}_{2}$ and other group-vi dichalcogenides}.
\newblock \emph{\bibinfo{journal}{Phys. Rev. Lett.}}
  \textbf{\bibinfo{volume}{108}}, \bibinfo{pages}{196802}
  (\bibinfo{year}{2012}).

\bibitem{Xi2015a}
\bibinfo{author}{Xi, X.} \emph{et~al.}
\newblock \bibinfo{title}{Ising pairing in superconducting {$\mathrm{NbSe_2}$}
  atomic~layers}.
\newblock \emph{\bibinfo{journal}{Nat. Phys.}} \textbf{\bibinfo{volume}{12}},
  \bibinfo{pages}{139--143} (\bibinfo{year}{2015}).

\bibitem{Sipos2008}
\bibinfo{author}{Sipos, B.} \emph{et~al.}
\newblock \bibinfo{title}{From {M}ott state to superconductivity in
  {$1{T}$-$\mathrm{TaS_2}$}}.
\newblock \emph{\bibinfo{journal}{Nat Mater}} \textbf{\bibinfo{volume}{7}},
  \bibinfo{pages}{960--965} (\bibinfo{year}{2008}).

\bibitem{Ritschel2015}
\bibinfo{author}{Ritschel, T.} \emph{et~al.}
\newblock \bibinfo{title}{Orbital textures and charge density waves in
  transition metal dichalcogenides}.
\newblock \emph{\bibinfo{journal}{Nat Phys}} \textbf{\bibinfo{volume}{11}},
  \bibinfo{pages}{328--331} (\bibinfo{year}{2015}).

\bibitem{Radisavljevic2011}
\bibinfo{author}{Radisavljevic, B.}, \bibinfo{author}{Radenovic, A.},
  \bibinfo{author}{Brivio, J.}, \bibinfo{author}{Giacometti, V.} \&
  \bibinfo{author}{Kis, A.}
\newblock \bibinfo{title}{Single-layer {$\mathrm{MoS_2}$} transistors}.
\newblock \emph{\bibinfo{journal}{Nat Nano}} \textbf{\bibinfo{volume}{6}},
  \bibinfo{pages}{147--150} (\bibinfo{year}{2011}).

\bibitem{Jariwala2014}
\bibinfo{author}{Jariwala, D.}, \bibinfo{author}{Sangwan, V.~K.},
  \bibinfo{author}{Lauhon, L.~J.}, \bibinfo{author}{Marks, T.~J.} \&
  \bibinfo{author}{Hersam, M.~C.}
\newblock \bibinfo{title}{Emerging device applications for semiconducting
  two-dimensional transition metal dichalcogenides}.
\newblock \emph{\bibinfo{journal}{ACS Nano}} \textbf{\bibinfo{volume}{8}},
  \bibinfo{pages}{1102--1120} (\bibinfo{year}{2014}).

\bibitem{Stojchevska2014}
\bibinfo{author}{Stojchevska, L.} \emph{et~al.}
\newblock \bibinfo{title}{Ultrafast switching to a stable hidden quantum state
  in an electronic crystal}.
\newblock \emph{\bibinfo{journal}{Science}} \textbf{\bibinfo{volume}{344}},
  \bibinfo{pages}{177--180} (\bibinfo{year}{2014}).

\bibitem{Yoshida2015}
\bibinfo{author}{Yoshida, M.}, \bibinfo{author}{Suzuki, R.},
  \bibinfo{author}{Zhang, Y.}, \bibinfo{author}{Nakano, M.} \&
  \bibinfo{author}{Iwasa, Y.}
\newblock \bibinfo{title}{Memristive phase switching in two-dimensional
  {1T}-{$\mathrm{TaS_2}$} crystals}.
\newblock \emph{\bibinfo{journal}{Sci. Adv.}} \textbf{\bibinfo{volume}{1}},
  \bibinfo{pages}{e1500606} (\bibinfo{year}{2015}).

\bibitem{Wilson1975a}
\bibinfo{author}{Wilson, J.}, \bibinfo{author}{Di~Salvo, F.} \&
  \bibinfo{author}{Mahajan, S.}
\newblock \bibinfo{title}{Charge-density waves and superlattices in the
  metallic layered transition metal dichalcogenides}.
\newblock \emph{\bibinfo{journal}{Advances in Physics}}
  \textbf{\bibinfo{volume}{24}}, \bibinfo{pages}{117--201}
  (\bibinfo{year}{1975}).

\bibitem{Vaskivskyi2015}
\bibinfo{author}{Vaskivskyi, I.} \emph{et~al.}
\newblock \bibinfo{title}{Controlling the metal-to-insulator relaxation of the
  metastable hidden quantum state in 1{T}-{$\mathrm{TaS_2}$}}.
\newblock \emph{\bibinfo{journal}{Science Advances}}
  \textbf{\bibinfo{volume}{1}} (\bibinfo{year}{2015}).

\bibitem{Hollander2015}
\bibinfo{author}{Hollander, M.~J.} \emph{et~al.}
\newblock \bibinfo{title}{Electrically driven reversible insulator--metal phase
  transition in 1{T}-{$\mathrm{TaS_2}$}}.
\newblock \emph{\bibinfo{journal}{Nano Letters}} \textbf{\bibinfo{volume}{15}},
  \bibinfo{pages}{1861--1866} (\bibinfo{year}{2015}).
\newblock \bibinfo{note}{PMID: 25626012}.

\bibitem{Guyader2017}
\bibinfo{author}{Guyader, L.~L.} \emph{et~al.}
\newblock \bibinfo{title}{Stacking order dynamics in the quasi-two-dimensional
  dichalcogenide {1T}-{$\mathrm{TaS_2}$} probed with {MeV} ultrafast electron
  diffraction}.
\newblock \emph{\bibinfo{journal}{Structural Dynamics}}
  \textbf{\bibinfo{volume}{4}}, \bibinfo{pages}{044020} (\bibinfo{year}{2017}).

\bibitem{Hovden2018}
\bibinfo{author}{Hovden, R.} \emph{et~al.}
\newblock \bibinfo{title}{Thickness and stacking sequence determination of
  exfoliated dichalcogenides ({1T}-{$\mathrm{TaS_2}$}, {2H}-{$\mathrm{MoS_2}$})
  using scanning transmission electron microscopy}.
\newblock \emph{\bibinfo{journal}{Microscopy and Microanalysis}}
  \textbf{\bibinfo{volume}{24}}, \bibinfo{pages}{387--395}
  (\bibinfo{year}{2018}).

\bibitem{Ma2016}
\bibinfo{author}{Ma, L.} \emph{et~al.}
\newblock \bibinfo{title}{A metallic mosaic phase and the origin of
  {M}ott-insulating state in {1T}-{$\mathrm{TaS_2}$}}.
\newblock \emph{\bibinfo{journal}{Nat Commun}} \textbf{\bibinfo{volume}{7}},
  \bibinfo{pages}{10956} (\bibinfo{year}{2016}).

\bibitem{Cho2016}
\bibinfo{author}{Cho, D.} \emph{et~al.}
\newblock \bibinfo{title}{Nanoscale manipulation of the {M}ott insulating state
  coupled to charge order in {1T}-{$\mathrm{TaS_2}$}}.
\newblock \emph{\bibinfo{journal}{Nat Commun}} \textbf{\bibinfo{volume}{7}},
  \bibinfo{pages}{10453} (\bibinfo{year}{2016}).

\bibitem{Cho2017}
\bibinfo{author}{Cho, D.} \emph{et~al.}
\newblock \bibinfo{title}{Correlated electronic states at domain walls of a
  mott-charge-density-wave insulator 1{T}-{$\mathrm{TaS_2}$}}.
\newblock \emph{\bibinfo{journal}{Nat Commun}} \textbf{\bibinfo{volume}{8}}
  (\bibinfo{year}{2017}).

\bibitem{Gerasimenko2019a}
\bibinfo{author}{Gerasimenko, Y.~A.}, \bibinfo{author}{Karpov, P.},
  \bibinfo{author}{Vaskivskyi, I.}, \bibinfo{author}{Brazovskii, S.} \&
  \bibinfo{author}{Mihailovic, D.}
\newblock \bibinfo{title}{Intertwined chiral charge orders and topological
  stabilization of the light-induced state of a prototypical transition metal
  dichalcogenide}.
\newblock \emph{\bibinfo{journal}{npj Quantum Materials}}
  \textbf{\bibinfo{volume}{4}} (\bibinfo{year}{2019}).

\bibitem{Brouwer1980}
\bibinfo{author}{Brouwer, R.} \& \bibinfo{author}{Jellinek, F.}
\newblock \bibinfo{title}{The low-temperature superstructures of
  {$1{T}$-$\mathrm{TaSe_2}$} and {$2{H}$-$\mathrm{TaSe_2}$}}.
\newblock \emph{\bibinfo{journal}{Physica B+C}} \textbf{\bibinfo{volume}{99}},
  \bibinfo{pages}{51--55} (\bibinfo{year}{1980}).

\bibitem{Spijkerman1997}
\bibinfo{author}{Spijkerman, A.}, \bibinfo{author}{de~Boer, J.~L.},
  \bibinfo{author}{Meetsma, A.}, \bibinfo{author}{Wiegers, G.~A.} \&
  \bibinfo{author}{van Smaalen, S.}
\newblock \bibinfo{title}{X-ray crystal-structure refinement of the nearly
  commensurate phase of {$1{T}$-$\mathrm{TaS_2}$} in {$(3+2)$}-dimensional
  superspace}.
\newblock \emph{\bibinfo{journal}{Phys. Rev. B}} \textbf{\bibinfo{volume}{56}},
  \bibinfo{pages}{13757--13767} (\bibinfo{year}{1997}).

\bibitem{Walker1983}
\bibinfo{author}{Walker, M.~B.} \& \bibinfo{author}{Withers, R.~L.}
\newblock \bibinfo{title}{Stacking of charge-density waves in {$1{T}$}
  transition-metal dichalcogenides}.
\newblock \emph{\bibinfo{journal}{Phys. Rev. B}} \textbf{\bibinfo{volume}{28}},
  \bibinfo{pages}{2766--2774} (\bibinfo{year}{1983}).

\bibitem{Nakanishi1984}
\bibinfo{author}{Nakanishi, K.} \& \bibinfo{author}{Shiba, H.}
\newblock \bibinfo{title}{Theory of three-dimensional orderings of
  charge-density waves in {$1{T}$-$\mathrm{TaX_2}$ (X: S, Se)}}.
\newblock \emph{\bibinfo{journal}{Journal of the Physical Society of Japan}}
  \textbf{\bibinfo{volume}{53}}, \bibinfo{pages}{1103--1113}
  (\bibinfo{year}{1984}).

\bibitem{Tanda1984}
\bibinfo{author}{Tanda, S.}, \bibinfo{author}{Sambongi, T.},
  \bibinfo{author}{Tani, T.} \& \bibinfo{author}{Tanaka, S.}
\newblock \bibinfo{title}{X-ray study of charge density wave structure in
  {$1{T}$-$\mathrm{TaS_2}$}}.
\newblock \emph{\bibinfo{journal}{Journal of the Physical Society of Japan}}
  \textbf{\bibinfo{volume}{53}}, \bibinfo{pages}{476--479}
  (\bibinfo{year}{1984}).

\bibitem{Ritschel2018}
\bibinfo{author}{Ritschel, T.}, \bibinfo{author}{Berger, H.} \&
  \bibinfo{author}{Geck, J.}
\newblock \bibinfo{title}{Stacking-driven gap formation in layered
  1{T}-{$\mathrm{TaS_2}$}}.
\newblock \emph{\bibinfo{journal}{Phys. Rev. B}} \textbf{\bibinfo{volume}{98}}
  (\bibinfo{year}{2018}).

\bibitem{Tosatti1976}
\bibinfo{author}{Tosatti, E.} \& \bibinfo{author}{Fazekas, P.}
\newblock \bibinfo{title}{On the nature of the low-temperature phase of
  {$1{T}$-$\mathrm{TaS_2}$}}.
\newblock \emph{\bibinfo{journal}{Le Journal de Physique Colloques}}
  \textbf{\bibinfo{volume}{37}}, \bibinfo{pages}{C4--165--C4--168}
  (\bibinfo{year}{1976}).

\bibitem{Ge2010}
\bibinfo{author}{Ge, Y.} \& \bibinfo{author}{Liu, A.~Y.}
\newblock \bibinfo{title}{First-principles investigation of the
  charge-density-wave instability in {$1{T}$-$\mathrm{TaSe_2}$}}.
\newblock \emph{\bibinfo{journal}{Phys. Rev. B}} \textbf{\bibinfo{volume}{82}},
  \bibinfo{pages}{155133} (\bibinfo{year}{2010}).

\bibitem{Lee2019}
\bibinfo{author}{Lee, S.-H.}, \bibinfo{author}{Goh, J.~S.} \&
  \bibinfo{author}{Cho, D.}
\newblock \bibinfo{title}{Origin of the insulating phase and first-order
  metal-insulator transition in {1T}-{$\mathrm{TaS_2}$}}.
\newblock \emph{\bibinfo{journal}{Physical Review Letters}}
  \textbf{\bibinfo{volume}{122}} (\bibinfo{year}{2019}).

\bibitem{Nakanishi1978a}
\bibinfo{author}{Nakanishi, K.} \& \bibinfo{author}{Shiba, H.}
\newblock \bibinfo{title}{Domain-like incommensurate charge-density-wave states
  and the first-order incommensurate-commensurate transitions in layered
  tantalum dichalcogenides. ii. 2{H}-polytype}.
\newblock \emph{\bibinfo{journal}{Journal of the Physical Society of Japan}}
  \textbf{\bibinfo{volume}{44}}, \bibinfo{pages}{1465--1473}
  (\bibinfo{year}{1978}).

\bibitem{Wu1989}
\bibinfo{author}{Wu, X.~L.} \& \bibinfo{author}{Lieber, C.~M.}
\newblock \bibinfo{title}{Hexagonal domain-like charge density wave phase of
  {$1{T}$-$\mathrm{TaS_2}$} determined by scanning tunneling microscopy}.
\newblock \emph{\bibinfo{journal}{Science}} \textbf{\bibinfo{volume}{243}},
  \bibinfo{pages}{4899} (\bibinfo{year}{1989}).

\bibitem{Park2019}
\bibinfo{author}{Park, J.~W.}, \bibinfo{author}{Cho, G.~Y.},
  \bibinfo{author}{Lee, J.} \& \bibinfo{author}{Yeom, H.~W.}
\newblock \bibinfo{title}{Emergent honeycomb network of topological excitations
  in correlated charge density wave}.
\newblock \emph{\bibinfo{journal}{Nat Commun}} \textbf{\bibinfo{volume}{10}}
  (\bibinfo{year}{2019}).

\bibitem{Ritschel2013}
\bibinfo{author}{Ritschel, T.} \emph{et~al.}
\newblock \bibinfo{title}{Pressure dependence of the charge density wave in
  {$1{T}$-$\mathrm{TaS_2}$} and its relation to superconductivity}.
\newblock \emph{\bibinfo{journal}{Phys. Rev. B}} \textbf{\bibinfo{volume}{87}},
  \bibinfo{pages}{125135} (\bibinfo{year}{2013}).

\bibitem{Han2015}
\bibinfo{author}{Han, T.-R.~T.} \emph{et~al.}
\newblock \bibinfo{title}{Exploration of metastability and hidden phases in
  correlated electron crystals visualized by femtosecond optical doping and
  electron crystallography}.
\newblock \emph{\bibinfo{journal}{Science Advances}}
  \textbf{\bibinfo{volume}{1}}, \bibinfo{pages}{e1400173--e1400173}
  (\bibinfo{year}{2015}).

\bibitem{Perfetti2008}
\bibinfo{author}{Perfetti, L.} \emph{et~al.}
\newblock \bibinfo{title}{Femtosecond dynamics of electronic states in the
  {M}ott insulator {$1{T}$-$\mathrm{TaS_2}$} by time resolved photoelectron
  spectroscopy}.
\newblock \emph{\bibinfo{journal}{New J. Phys.}} \textbf{\bibinfo{volume}{10}},
  \bibinfo{pages}{053019 (17pp)} (\bibinfo{year}{2008}).

\bibitem{Hellmann2010a}
\bibinfo{author}{Hellmann, S.} \emph{et~al.}
\newblock \bibinfo{title}{Ultrafast melting of a charge-density wave in the
  {M}ott insulator {{$1{T}$}-$\mathrm{TaS_2}$}}.
\newblock \emph{\bibinfo{journal}{Phys. Rev. Lett.}}
  \textbf{\bibinfo{volume}{105}}, \bibinfo{pages}{187401}
  (\bibinfo{year}{2010}).

\bibitem{Hellmann2012}
\bibinfo{author}{Hellmann, S.} \emph{et~al.}
\newblock \bibinfo{title}{Time-domain classification of charge-density-wave
  insulators}.
\newblock \emph{\bibinfo{journal}{Nat Commun}} \textbf{\bibinfo{volume}{3}},
  \bibinfo{pages}{1069--} (\bibinfo{year}{2012}).

\bibitem{Ligges2018}
\bibinfo{author}{Ligges, M.} \emph{et~al.}
\newblock \bibinfo{title}{Ultrafast doublon dynamics in photoexcited
  {1T}-{$\mathrm{TaS_2}$}}.
\newblock \emph{\bibinfo{journal}{Physical Review Letters}}
  \textbf{\bibinfo{volume}{120}} (\bibinfo{year}{2018}).

\bibitem{CrysAlisPRO}
\bibinfo{title}{{Oxford Diffraction /Agilent Technologies UK Ltd{,} Yarnton{,}
  England}}.

\end{thebibliography}
\end{document}